\documentclass[a4paper]{spie}  

\usepackage[]{graphicx}
\usepackage{amssymb,amsmath,psfrag,url}

\newcommand{\grad}{\ensuremath{^\circ}}

\title{Rigorous FEM-Simulation of EUV-Masks: Influence of Shape and Material
Parameters} 

\author{
Jan Pomplun\supit{\,ab},
Sven Burger\supit{\,ab}, 
Frank Schmidt\supit{\,ab}, 
Lin Zschiedrich\supit{\,ab},
Frank Scholze\supit{\,c},
Christian Laubis\supit{\,c},
Uwe Dersch\supit{\,d},
\skiplinehalf
\supit{a}
Zuse Institute Berlin,
Takustra{\ss}e 7,
D\,--\,14\,195 Berlin,
Germany\\
DFG Forschungszentrum {\sc Matheon},
Stra{\ss}e des 17.\,Juni 136, 
D\,--\,10\,623 Berlin,
Germany
\smallskip\\
\supit{b}
JCMwave GmbH,
Haarer Stra{\ss}e 14a,
D\,--\,85\,640 Putzbrunn, 
Germany
\smallskip\\
\supit{c}
Physikalisch-Technische Bundesanstalt,
EUV radiometry\\
Abbestra{\ss}e 2\,--\,12,
D\,--\,10\,587 Berlin,
Germany
\smallskip\\
\supit{d}
Advanced Mask Technology Center GmbH \& Co. KG\\
R{\"a}hnitzer Allee 9,
D\,--\,01\,109 Dresden,
Germany
}

\authorinfo{
Corresponding author: J. Pomplun\\
URL: http://www.zib.de/nano-optics/\\
Email: pomplun@zib.de
}

\begin{document} 
  \maketitle 

Copyright 2006  Society of Photo-Optical Instrumentation Engineers.\\
This paper has been published in Proc.~SPIE {\bf 6349}, 63493D
(2006),  
({\it 26th Annual BACUS Symposium on Photomask Technology, 
P.~M.~Martin, R. J. Naber, Eds.})
and is made available 
as an electronic reprint with permission of SPIE. 
One print or electronic copy may be made for personal use only. 
Systematic or multiple reproduction, distribution to multiple 
locations via electronic or other means, duplication of any 
material in this paper for a fee or for commercial purposes, 
or modification of the content of the paper are prohibited.

\begin{abstract}
We present rigorous simulations of EUV masks with technological imperfections like side-wall angles and corner roundings. We perform an optimization of two different geometrical parameters in order to fit the numerical results to results obtained from experimental scatterometry measurements.
For the numerical simulations we use an adaptive finite element approach on irregular meshes \cite{Burger2005w}. This gives us the opportunity to model geometrical structures accurately. Moreover we comment on the use of domain decomposition techniques for EUV mask simulations \cite{Zschiedrich2005b}.
Geometric mask parameters have a great influence on the diffraction pattern. We show that using accurate simulation tools it is possible to deduce the relevant geometrical parameters of EUV masks from scatterometry measurements.

This work results from a collaboration between AMTC (mask fabrication), Physikalisch-Technische Bundes\-anstalt (scatterometry) and ZIB/JCMwave (numerical simulation).
\end{abstract}


\section{Introduction}
Extreme ultraviolet (EUV) lithography is considered as the main candidate for further miniaturization of computer technology. Since compared to state-of-the art photomasks, EUV masks are illuminated at oblique incidence, the quality of pattern profiles becomes important due to shadowing effects \cite{Sugawara05a,Sugawara05b}. Consequently, there is a need for adequate destruction free pattern profile metrology techniques, allowing characterization of mask features down to a typical size of $100\,$nm.

Here we present an indirect method for the determination of geometrical EUV mask parameters \cite{SCHOLZE1}. Experimental scatterometry measurements are compared to numerical simulations of EUV masks using the finite element method (FEM).

\section{Characterization of EUV masks by EUV scatterometry}
\label{sec::scatterometry}
\begin{figure}
(a)\hspace{9cm}(b)\\\vspace{0.4cm}\\
\psfrag{euv}{light source}
\psfrag{mult}{Mo/Si multilayer}
\psfrag{det}{detector with slit}
\psfrag{abs}{absorber stack}
\psfrag{Si}{Si-cap}
\psfrag{cd}{CD}
\includegraphics[width=7cm]{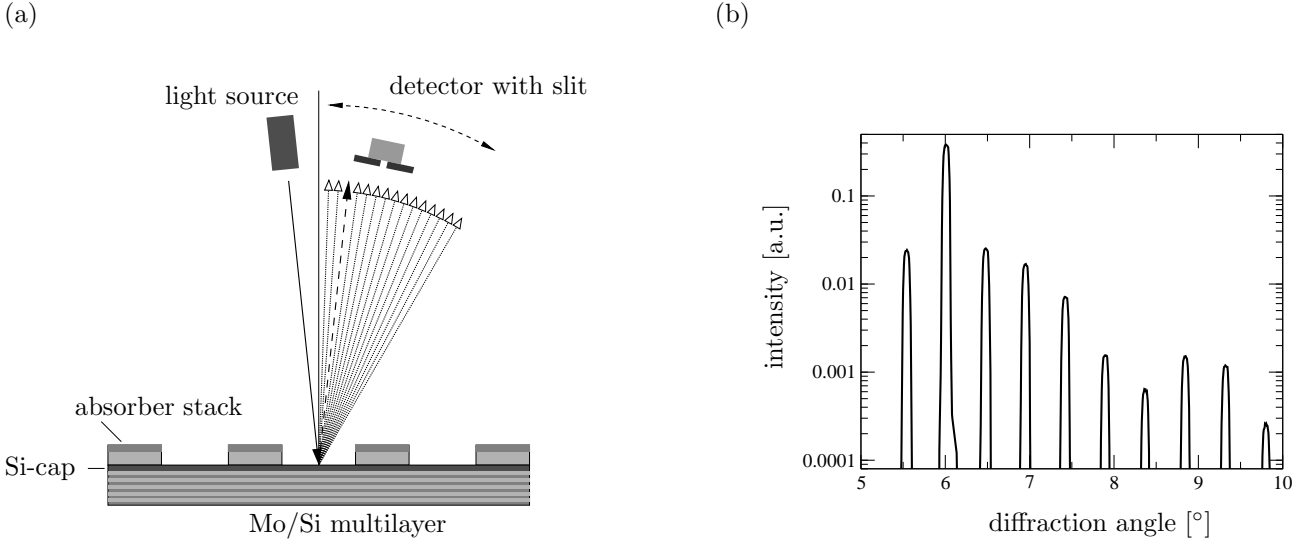}
\hfill
\psfrag{diff}{diffraction angle [\grad]}
\psfrag{intensity}{intensity [a.u.]}
\includegraphics[width=7cm]{fig/h4scatter.eps}
\caption{\label{fig:h4scatter} (a) Experimental setup for scatterometry experiment with fixed incident angle of $6\grad$ and variable angle of detection $\theta_{out}$; (b) Result of single wavelength scatterometry measurement at $\lambda=13.65\,$nm. Diffraction orders appear as peaks with finite width, the zeroth diffraction peak is centered around $6\grad$.}
\end{figure}

Single wavelength scatterometry, the analysis of light diffracted from a periodic structure, is a well suited tool for analysis of the geometry of EUV masks. Since scatterometry only needs a light source and a simple detector with no imaging lens system, its setup is inexpensive and offers no additional technical challenges. Fig. \ref{fig:h4scatter}(a) shows a sketch of the experimental setup. Light of fixed wavelength and fixed incident angle is reflected from the mask and the intensity of the reflected light is measured in dependence on the diffraction angle.
The use of EUV light for mask characterization is advantageous because it fits the small feature sizes on EUV masks. Diffraction phenomena are minimized, and of course the appropriate wavelength of the resonant structure of the underlying multilayer is chosen. Light is not only reflected at the top interface of the mask but all layers in the stack contribute to reflection. Therefore one expects that EUV radiation provides much more information on relevant EUV mask features than conventional long wavelength methods.

All measurements for the present work were performed by the Physikalisch-Technische Bundesanstalt (PTB) at the electron storage ring BESSY II \cite{Ulm98}. PTB's EUV reflectometer installed in the soft X-ray radiometry beamline allows high-accuracy measurements of very large samples with diameters up to $550\,$mm \cite{Tuemmler03,Scholze05,Scholze03}.

\section{FEM simulation of EUV scatterometry}
\label{sec::fem}

\begin{figure}[ht]
\centering
\includegraphics[width=8cm]{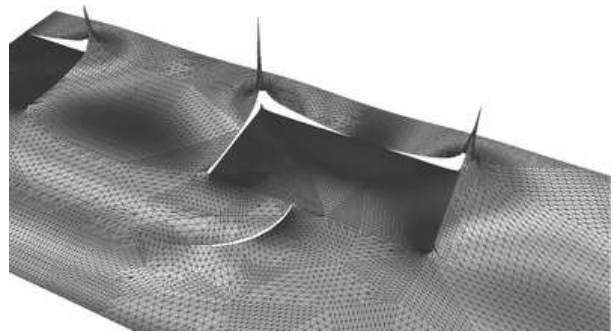}\\
\caption{\label{fig:undercut}FEM solution for the electric field propagating through a phase mask. The electric field has singular behaviour at corners of the absorber and discontinuities at material interfaces.}
\end{figure}

Fig. \ref{fig:h4scatter}(b) shows the result of a scatterometry measurement of an EUV test mask (see table \ref{table:mask}) considered in the present work.
\begin{table}[h]
\begin{centering}
\begin{tabular}{|l|c|}
\hline
Stack & Testmask \\
\hline 
\hline
ARC + TaN-Absorber& $67\,$nm\\
 SiO$_{2}$-Buffer & $10\,$nm\\
Si-Capping layer& $11\,$nm\\
Multilayer & Mo/Si \\
\hline
\end{tabular}
\hspace{3cm}
\begin{tabular}{|l|c|c}
\hline
Design parameter & Testmask \\
\hline 
\hline 
Absorber stack sidewall angle $\alpha$& 90\grad\\
Pitch  & 840nm\\
Top critical dimension & 140nm \\
\hline
\end{tabular}
\end{centering}
\caption{\label{table:mask}Design parameters (see also Fig. \ref{fig:defineParam}) for EUV test mask produced by AMTC.}
\end{table}
The position of the diffraction angles provide information about the pitch of the EUV absorber pattern. However the intensities of the diffraction orders do not carry direct information about other topological features of the mask. The determination of these features from a scatterometry measurement is a challenging inverse problem and a hot topic of actual research. Accurate and fast numerical simulation of the scattering experiment thereby plays a vital role. The FEM method is particularly suited for this application.
\begin{figure}[ht]
\includegraphics[width=4cm,height=2.63cm]{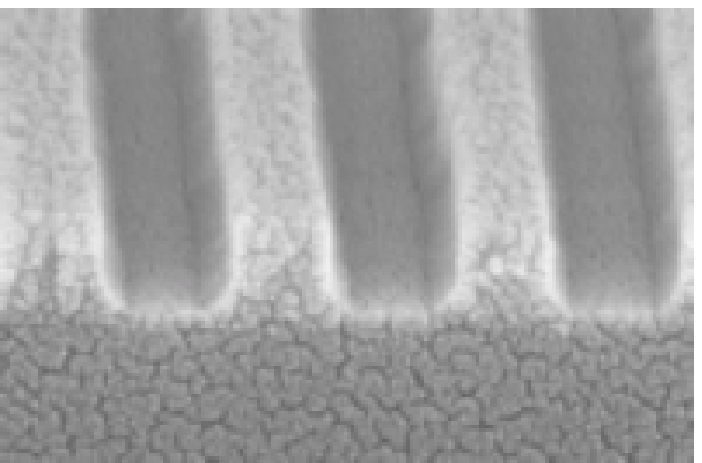}\hfill
\includegraphics[width=4cm,height=2.63cm]{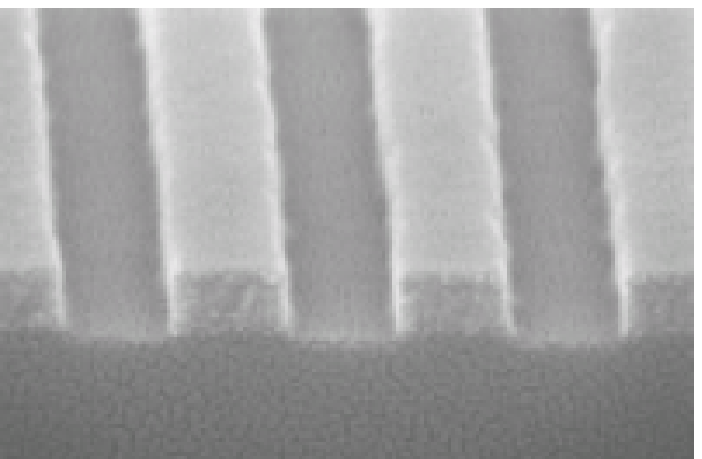}\hfill
\includegraphics[width=4cm,height=2.63cm]{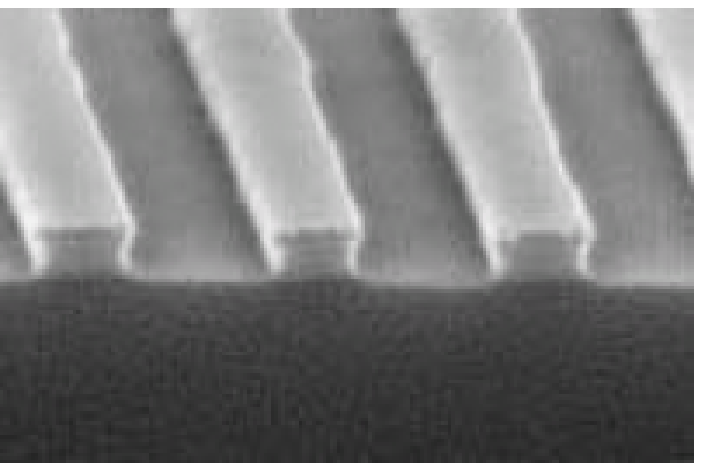}\hfill
\includegraphics[width=4cm,height=2.63cm]{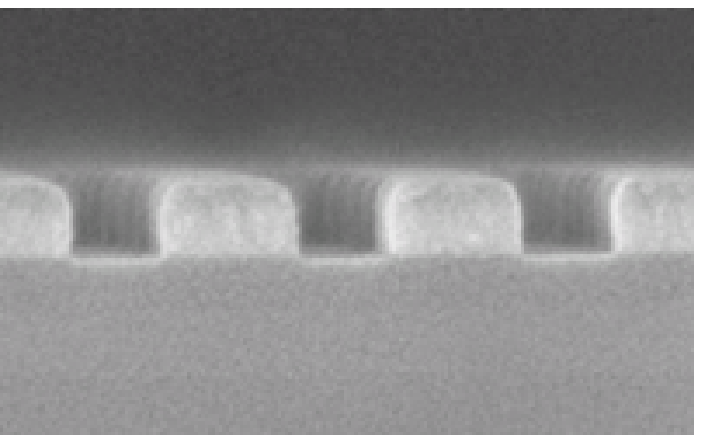}\\
\includegraphics[width=4cm,height=2.63cm]{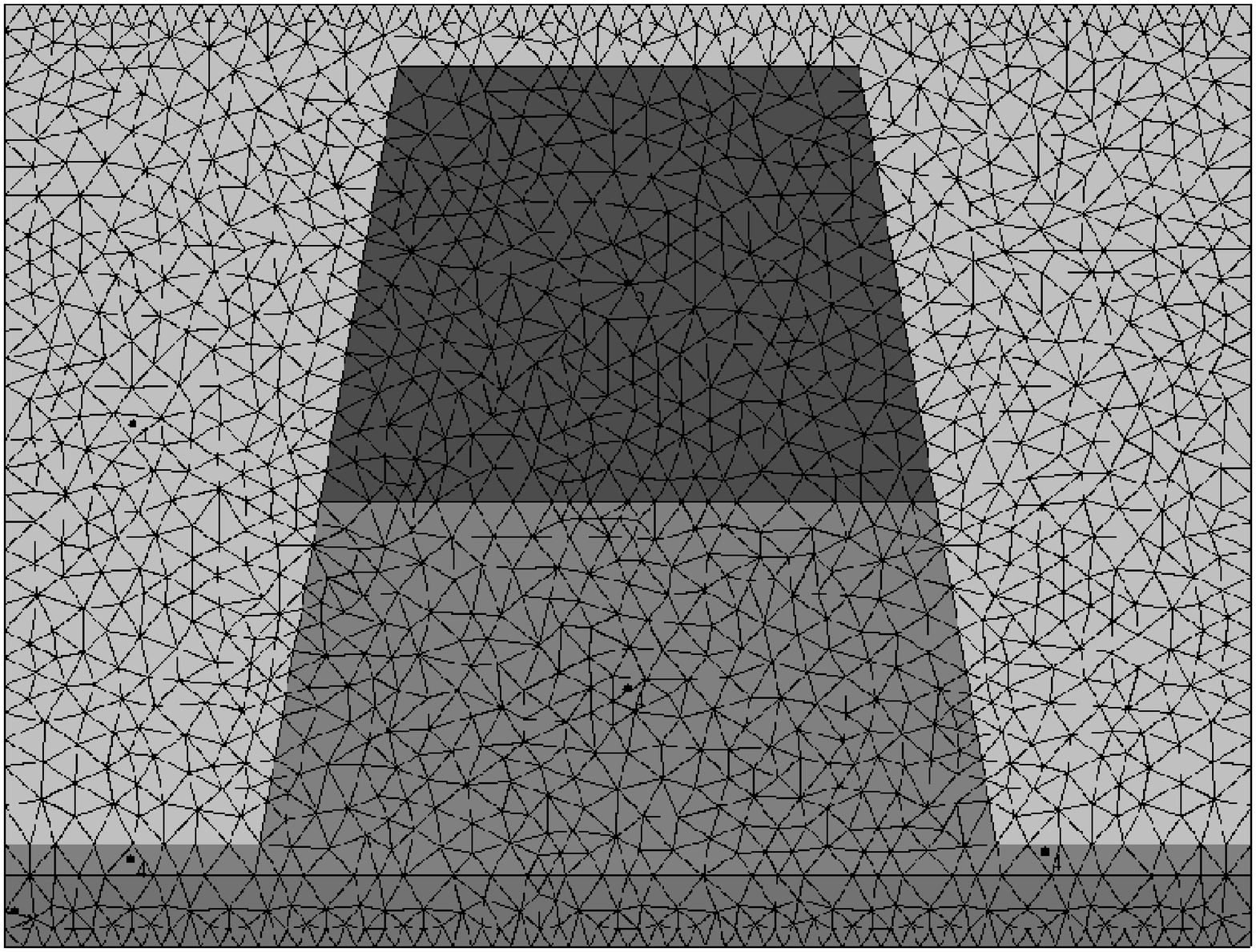}\hfill
\includegraphics[width=4cm,height=2.63cm]{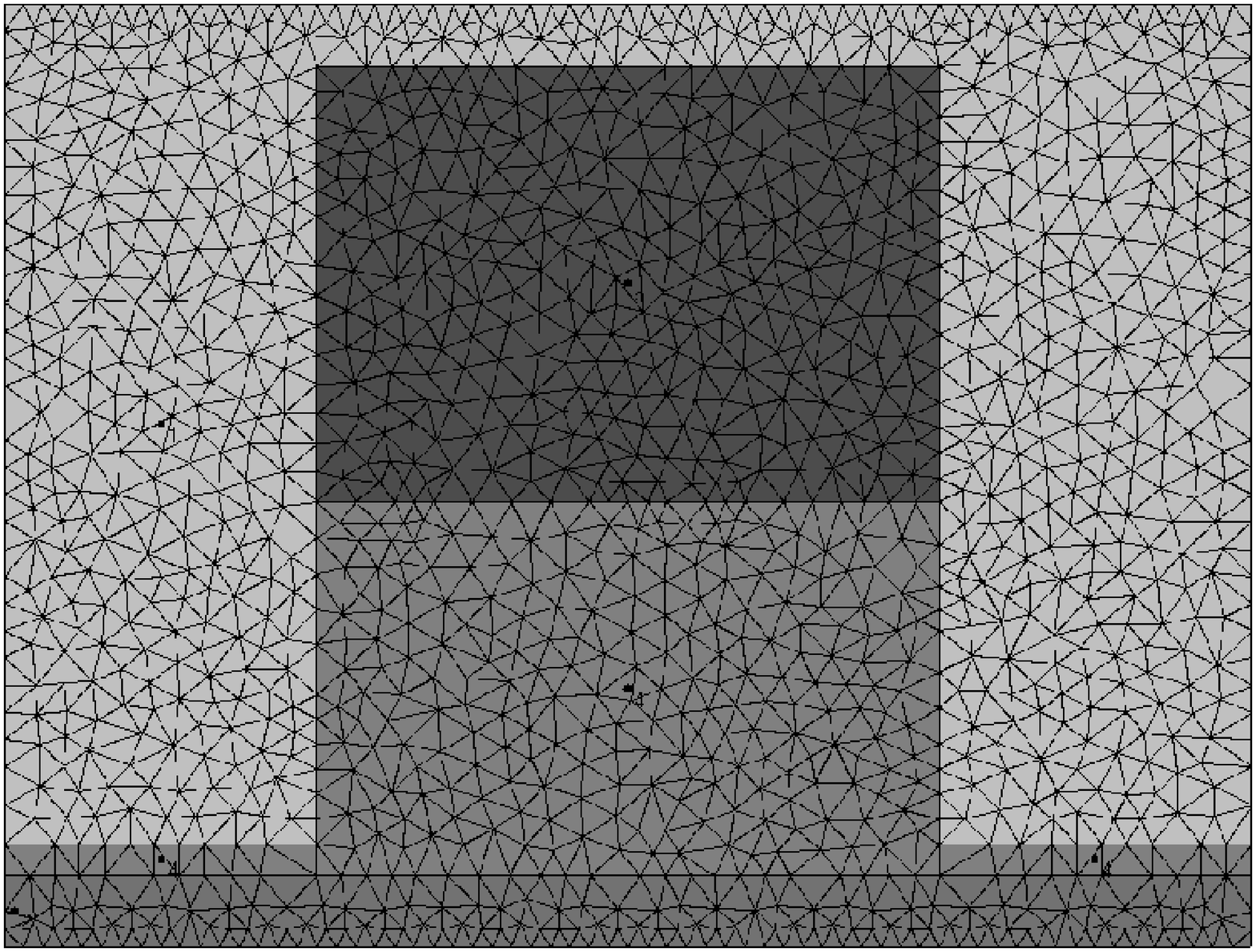}\hfill
\includegraphics[width=4cm,height=2.63cm]{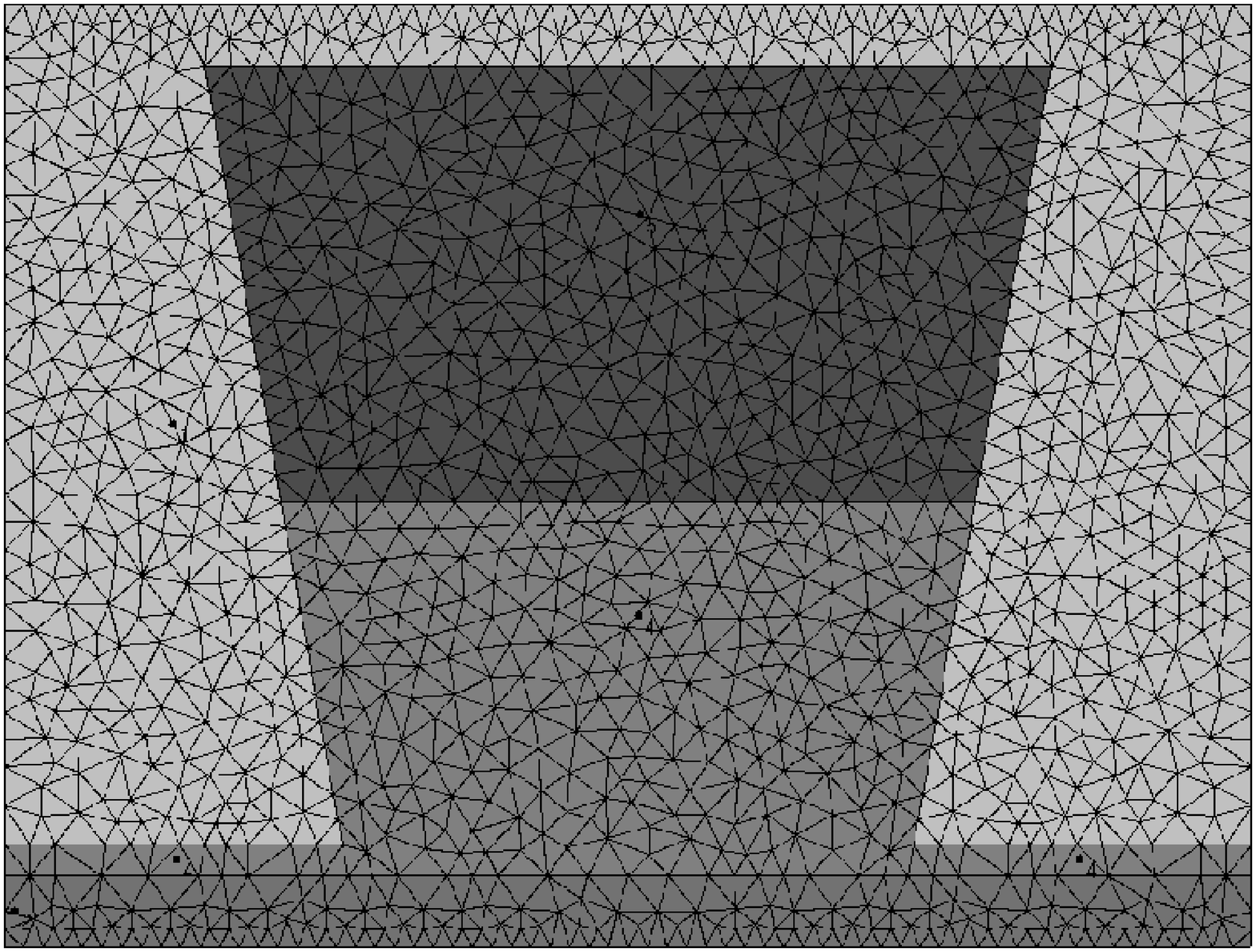}\hfill
\includegraphics[width=4cm,height=2.63cm]{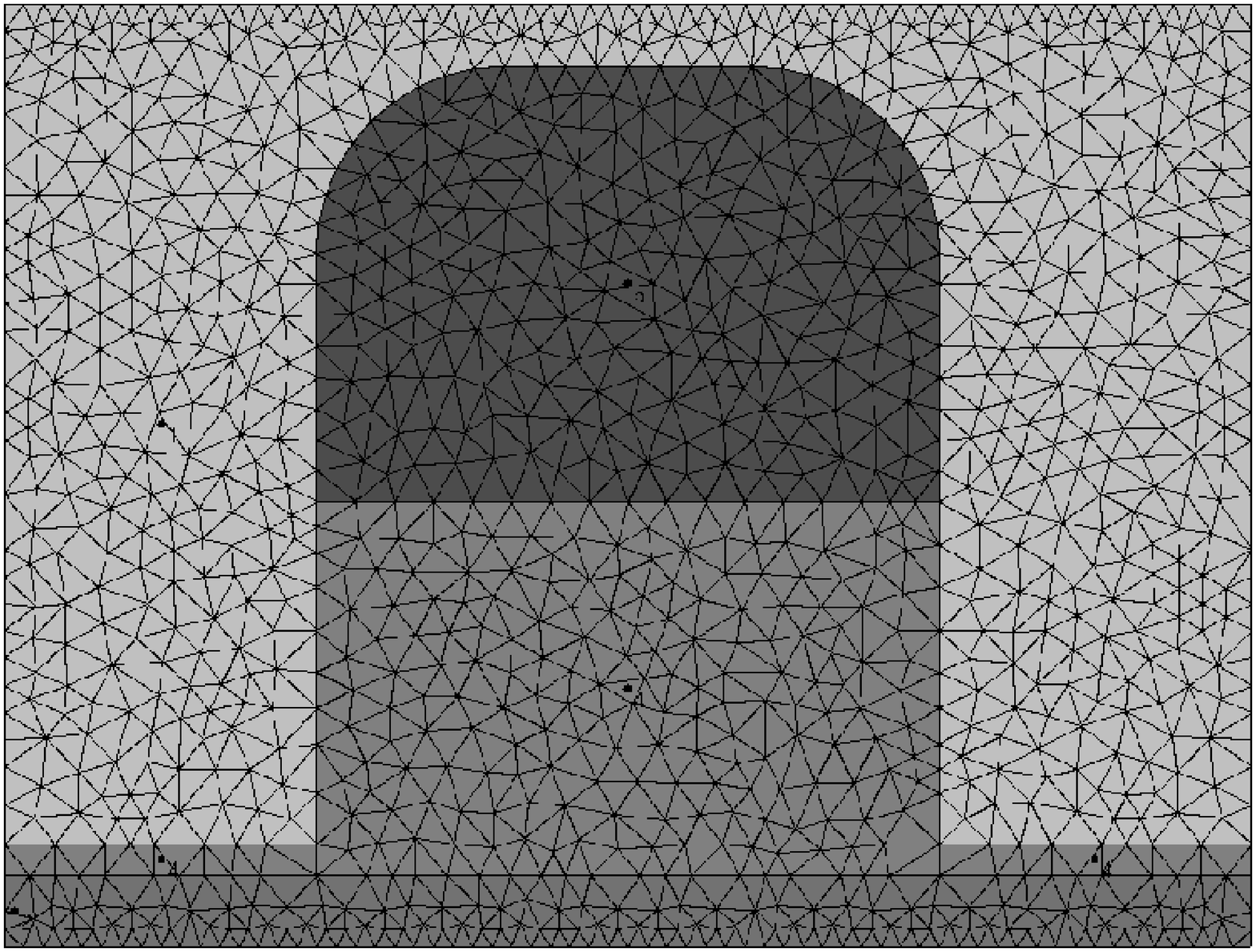}\\
\caption{\label{fig:triang}SEM pictures of EUV mask patterns and corresponding triangulated geometries for FEM computation.}
\end{figure}
It has several advantages \cite{Burger2005bacus}:
\begin{itemize}
\item
Maxwell's equations describing the scattering problem are solved rigorously without approximations.
\item 
The flexibility of triangulations allows modeling of virtually arbitrary structures, as illustrated in Fig. \ref{fig:triang}.
\item 
Adaptive mesh-refinement strategies lead to very accurate results and small computational times which are crucial points for application of a numerical method to the characterization of EUV masks.
\item
Choosing appropriate localized ansatz functions for the solution of Maxwell's equations physical properties of the electric field like discontinuities or singularities can be modeled very accurately and don't give rise to numerical problems, see Fig.\ref{fig:undercut}.
\item
It is mathematically proven that the FEM approach converges with a fixed convergence rate towards the exact solution of Maxwell-type problems for decreasing mesh width of the triangulation. Therefore it is easy to check if numerical results can be trusted.
\end{itemize}
Throughout this paper we use the FEM solver JCMharmony for numerical solution of Maxwell's equations. JCMharmony has been successfully applied to a wide range of electromagnetic field computations including waveguide structures \cite{Burger2005a}, DUV phase masks \cite{Burger2005bacus}, and other nano-structured materials \cite{Enkrich2005a,Kalkbrenner2005a}. It provides higher order edge elements, multigrid methods, a-posteriori error control, adaptive mesh refinement, etc. Furthermore a special domain decomposition algorithm implemented in JCMharmony is utilized for simulation of EUV masks \cite{Zschiedrich2005b}. Light propagation in the multilayer stack beneath the absorber pattern can be determined analytically. The domain decomposition algorithm combines the analytical solution of the multilayer stack with the FEM solution of the absorber pattern, dramatically decreasing computational time and increasing accuracy of simulation results.

\section{Determination of EUV mask parameters with FEM simulation}
\begin{figure}[h]
\psfrag{alpha}{$\alpha$}
\psfrag{R}{R}
\psfrag{cd}{topCD}
\hspace{2cm}(a)\hspace{8cm}(b)
\begin{center}
\includegraphics[width=5cm]{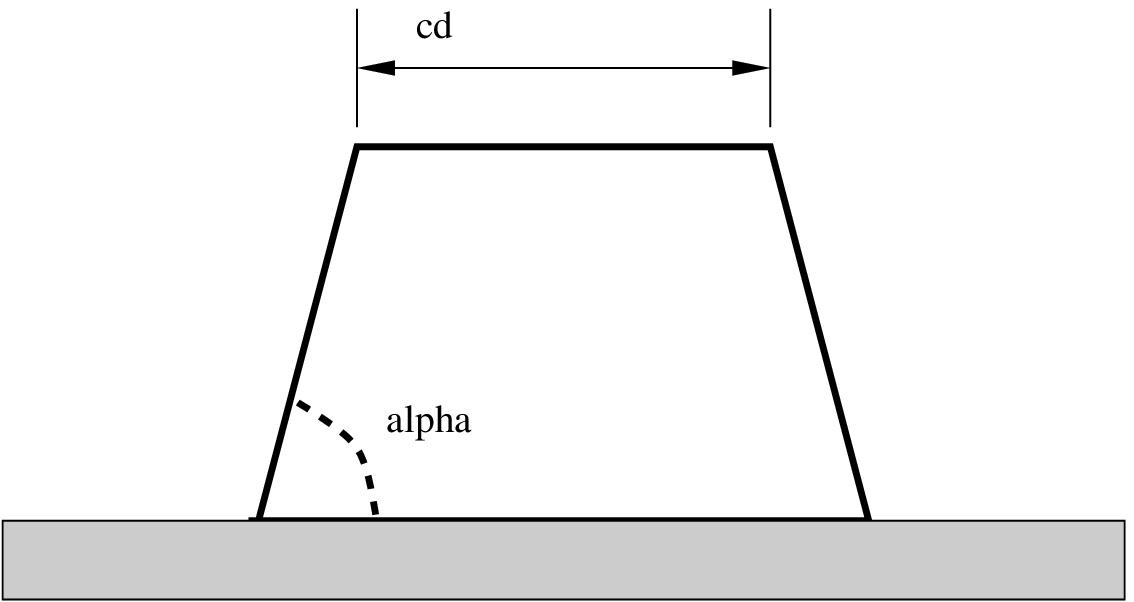}\hspace{3cm}
\includegraphics[width=5cm]{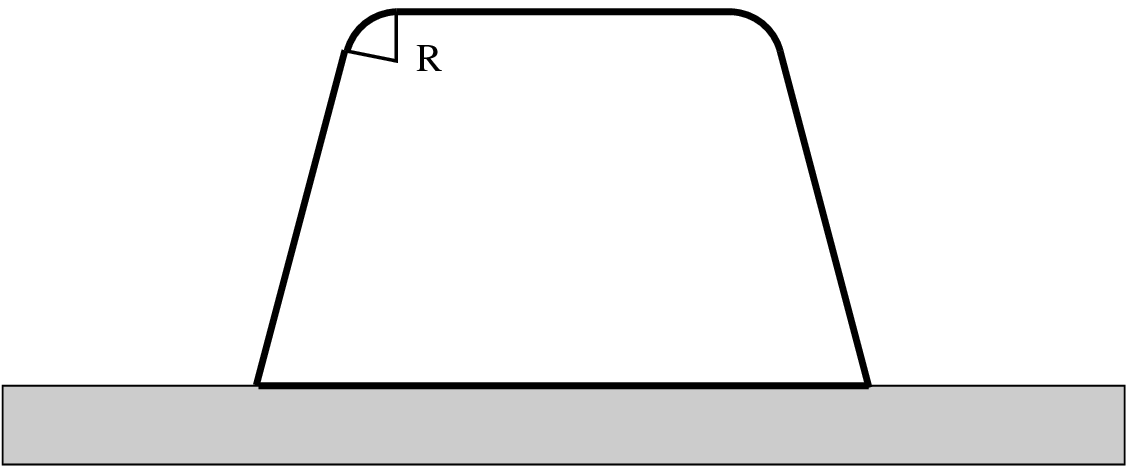}\hfill
\end{center}
\caption{\label{fig:defineParam}Parameters of EUV mask pattern: (a) Absorber stack sidewall angle $\alpha$ and top critical dimension topCD; (b) absorber edge Radius R}
\end{figure}
The idea of characterising EUV masks with scatterometry measurements and FEM simulations is the comparison of the intensities $I^{\mbox{exp}}_{n}$ of experimental diffraction orders $n$ with numerically obtained results \cite{SCHOLZE1}. First the geometry of the EUV mask is modeled using a finite number of parameters. Then scattering from the EUV mask is simulated and the intensities $I^{\mbox{sim}}_{n}$ of the diffraction orders are computed. The deviation $\xi$ between experimental and numerical intensities is computed and the parameters which minimize this deviation determined. This leads to a finite dimensional optimization problem.
In the experimental setup light of each diffraction order is always reflected into a finite solid angle leading to peaks with finite width, see Fig. \ref{fig:h4scatter}(b). The experimental intensities $I^{\mbox{exp}}_{n}$ that are used are given as the heights of these peaks which are only proportional to the whole intensity diffracted into an order (i.e. the integral over a peak). Numerically we determine the whole intensity of each diffraction order and therefore we have to scale the simulated intensities $I^{\mbox{sim}}_{n}$ uniformly with a factor $\gamma$ before determining the deviation $\xi$:
\begin{equation}
\xi^{2}=\sum_{n}\left(\frac{\gamma I^{\mbox{sim}}_{n}-I^{\mbox{exp}}_{n}}{I^{\mbox{exp}}_{n}}\right)^{2},\label{error}
\end{equation}
where the global scaling factor $\gamma$ is determined by minimizing $\xi^{2}$ with respect to $\gamma$:
\begin{equation}
\partial_{\gamma}\left(\xi^{2}\right)=0\quad\Leftrightarrow\quad
\gamma=\frac{\sum\limits_{n}\left(\frac{I_{n}^{\mbox{sim}}}{I_{n}^{\mbox{exp}}}\right)^{2}}{\sum\limits_{n}\frac{I_{n}^{\mbox{sim}}}{I_{n}^{\mbox{exp}}}}.\label{gamma}
\end{equation}
\vspace{0.5cm}
\begin{figure}[h]
\centering
\psfrag{ref}{$R$}
\psfrag{lambda}{$\lambda$ [nm]}
\psfrag{134}{$\lambda_{1}$}
\psfrag{1365}{$\lambda_{2}$}
\psfrag{139}{$\lambda_{3}$}
\includegraphics[width=6cm]{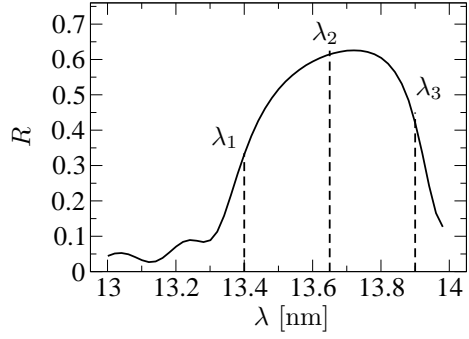}\hfill
\caption{\label{fig:h4bright}Brightfield measurement of multilayer of AMTC test mask: Reflectivity $R$ of open multilayer in dependence on incident wavelength for fixed incident angle of $6\grad$. Determination of EUV mask geometry was performed at three spectral wavelengths $\lambda_{1}=13.4\,$nm, $\lambda_{2}=13.65\,$nm, $\lambda_{3}=13.9\,$nm.  }
\end{figure}

The described procedure was applied to an EUV test mask produced by AMTC. The unknown mask parameters of interest were the sidewall angle of the absorber stack (which was restricted to $\alpha\le 90\grad$) and the top critical dimension (defined in Fig. \ref{fig:defineParam}).

The search for the optimal parameter set of the layout was performed using the Nelder-Mead simplex algorithm. As starting point the AMTC design parameters were chosen. In order to evaluate the results of our procedure scatterometry measurements and FEM simulations were compared at three different EUV wavelengths, shown in Fig. \ref{fig:h4bright}.

\begin{figure}
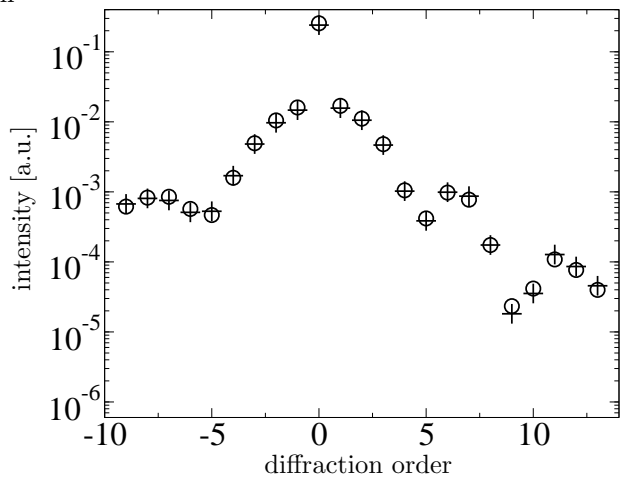
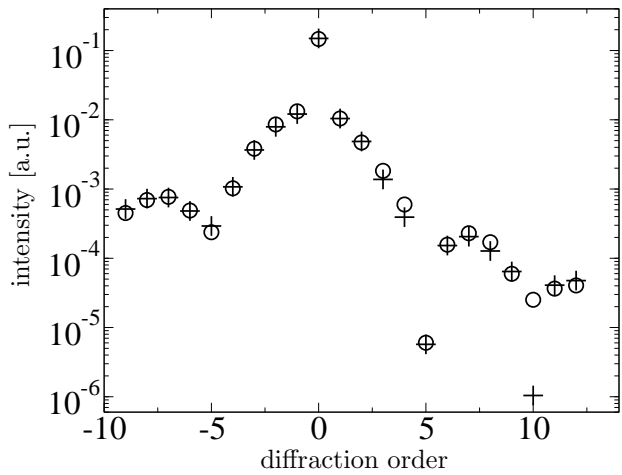

\psfrag{diffraction}{diffraction order}
\psfrag{int}{intensity [a.u.]}
\psfrag{exp}{experimental}
\psfrag{sim}{FEM computation}
$\lambda_{1}=13.4\,$nm\\\phantom{.......................}\includegraphics[width=8cm]{fig/scatt134.eps}\\
$\lambda_{2}=13.65\,$nm\\\phantom{.......................}\includegraphics[width=8cm]{fig/scatt1365.eps}\\
$\lambda_{3}=13.9\,$nm\\\phantom{.......................}\includegraphics[width=8cm]{fig/scatt139.eps}
\caption{\label{fig:scatt}Comparison between experimental scatterometry measurement and FEM computation of diffraction orders for different wavelengths $\lambda_{i}$ of incident EUV light.}
\end{figure}

A comparison of diffraction orders obtained from scatterometry and FEM simulation is shown in Fig. \ref{fig:scatt} for the optimal EUV mask parameters found during optimization. We see excellent agreement at all wavelengths of incident EUV radiation. Only the tenth diffraction order at $\lambda_{3}=13.9\,$nm differs. Here the simulated intensity is much lower than the experimental and therefore the deviation $\xi^{2}$ much larger than for $\lambda_{1}=13.4\,$nm and $\lambda_{2}=13.65\,$nm, see table \ref{table:result}.
\begin{figure}
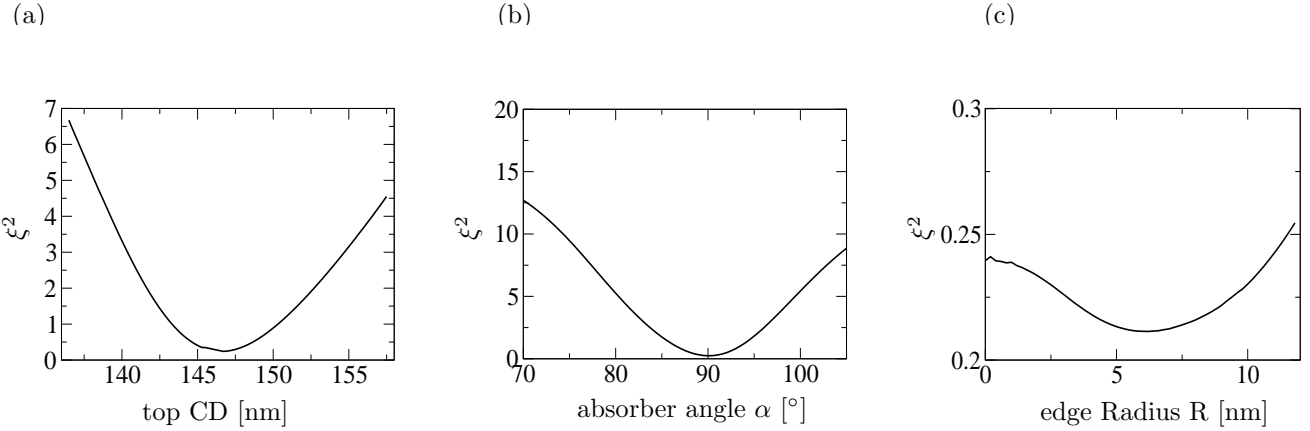

(a)\hspace{6cm}(b)\hspace{6cm}(c)\\\vspace{0.55cm}\\
\psfrag{xi}{$\xi^{2}$}
\psfrag{angle}{absorber angle $\alpha$ [\grad]}
\psfrag{width}{top CD [nm]}
\psfrag{radius}{edge Radius R [nm]}
\includegraphics[width=5.1cm,height=4.3cm]{fig/width.eps}\hfill
\includegraphics[width=5.1cm,height=4.3cm]{fig/angle.eps}\hfill
\includegraphics[width=5.1cm,height=4.3cm]{fig/corner.eps}\hfill
\caption{\label{fig:paramError}Dependence of deviation $\xi^{2}$ (see Eq. \ref{error}) on geometrical parameters of EUV mask. Fixed parameters: (a) $\alpha=90.0\grad$, $R=0\,$nm; (b) top CD=$146.5\,$nm, $R=0\,$nm; (c) $\alpha=90.0\grad$, top CD=$146.5\,$nm.}
\end{figure}
The geometrical parameters which belong to the best fitting simulations are shown in table \ref{table:result} in comparison to the desired design values for the mask.
\begin{table}[h]
\centering
\begin{tabular}{|l|c|c|c|c|}
\hline
geometrical parameter& design value & FEM $\lambda_{1}=13.4\,$nm& FEM $\lambda_{2}=13.65\,$nm& FEM $\lambda_{3}=13.9\,$nm\\
\hline
\hline
$\alpha$ [\grad]& 90 & 87.9 & 90.0 & 90.0\\
top CD [nm] & 140 & 145.7 & 146.5 & 146.5\\
\hline
\hline
deviation $\xi^{2}$ & & 0.18 & 0.24 & 1.32\\
\hline
\end{tabular}
\caption{\label{table:result}Deviation $\xi^{2}$ and geometrical parameters of EUV mask obtained from FEM computation at different wavelengths.}
\end{table}
We see that the best fitting geometries agree extremely well for $\lambda_{2}=13.65\,$nm and $\lambda_{3}=13.9\,$nm. For $\lambda_{1}=13.4\,$nm the absorber angle is $2\grad$ (2.2\%) smaller and the top CD is $0.8\,$nm (0.5\%) smaller than for $\lambda_{2}$ and $\lambda_{3}$.

Fig. \ref{fig:paramError}(a), (b) shows how the deviation $\xi^{2}$ between experimental and simulated diffraction orders depends on the top critical dimension and the absorber angle $\alpha$. It grows strongly with increasing distance from the optimal geometrical parameters. This shows that the presented method is very robust.

\begin{figure}
\centering
\psfrag{diff0}{diffraction order 0}
\psfrag{diff1}{diffraction order 1}
\psfrag{rel}{relative error}
\psfrag{number}{number of unknowns}
\includegraphics[width=7cm]{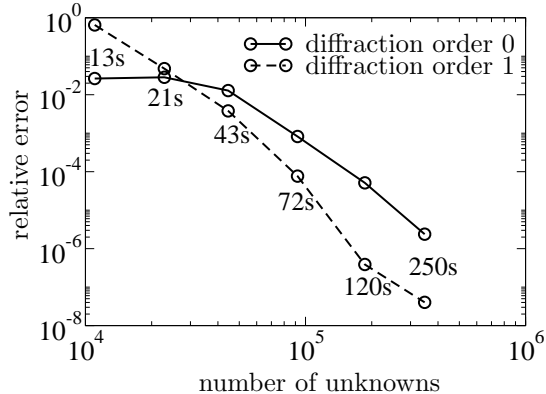}\hfill
\caption{\label{fig:conv}Convergence of FEM method: relative error of intensity of first two diffraction orders in dependence on number of unknowns of FEM computation.}
\end{figure}

As a further geometrical parameter the absorber edge radius $R$ was considered, see Fig. \ref{fig:defineParam}(b). The best fitting value for $R$ was determined at the incident wavelength $\lambda_{2}=13.65\,$nm and the optimal values for top CD = $146.5\,$nm and $\alpha=90.0\grad$. Fig. \ref{fig:paramError}(c) shows a minimal deviation between experiment and simulation for $R=6.2\,$nm. We see that the edge radius does not have such a great effect on the diffraction orders since the deviation $\xi^{2}$ hardly changes compared to the effects of the top CD (a) and absorber angle (b). This confirms further that for the determination of more sophisticated geometrical parameters very accurate simulations are crucial. As already mentioned the convergence of the FEM method is mathematically proven and it is therefore a very good choice for the presented method. Fig. \ref{fig:conv} shows the convergence of the zeroth and first diffraction order. We see the relative error in dependence on the number of unknowns of the FEM computation (i.e. a coarser triangulation). Furthermore the computational time on a standard PC ($3.4\,$GHz Intel Pentium 4, 1GB RAM) is shown at each refinement step of the grid. After $72\,$s we already have a relative error of $10^{-3}$ much smaller than the experimental uncertainty of about $0.01$. A short computation time also becomes crucial for the determination of mask parameters when choosing a larger number of independent geometrical parameters and performing the search for the optimal values in a higher dimensional space. We expect that scatterometry measurements at several different wavelengths will become very important for the presented method when characterizing EUV masks in greater detail. In order to further validate the geometrical parameters of the EUV mask obtained via scatterometry and FEM simulation a comparison to direct measurements like atomic force microscopy is planned. These measurements will be carried out at AMTC.
\vspace{1cm}


\section{Conclusions}
\label{sec::conclusions}
We demonstrated that single wavelength scatterometry in combination with FEM simulations is a promising candidate for an accurate and  robust destruction free characterization of EUV masks. Thereby experimental diffraction orders are compared to FEM simulations of EUV masks. For FEM simulations the EUV mask is first described with a finite number of geometrical parameters like sidewall angles, line widths, corner roundings, etc. and then the best fitting values determined by minimizing the deviation of experimental and numerical data.

Here we considered the top critical dimension, the sidewall angle and the edge radius of the absorber stack of an EUV mask as unknown geometrical parameters. The search for the best fitting geometry at three different wavelengths gave nearly the same values for the top critical dimension and the absorber sidewall angle proving both robustness and accuracy of the method. Furthermore the absorber edge radius had only minor influence on the numerical diffraction pattern.

We showed that very accurate numerical simulations are crucial for detailed geometric characterization of EUV masks using scatterometry data.
The FEM method is well suited for the simulation of EUV masks since it allows computation of nearly arbitrary geometries, is very accurate and very fast. Thereby very fast simulation of the EUV mask with a fixed parameter set provides a precondition for the solution of the given inverse problem.

\bibliography{myBib}
\bibliographystyle{spiebib}

\end{document}